\documentstyle[12pt]{article}

\input{pst-plot}

\input{pst-node}

\begin{document}

\begin{titlepage}

\hfill LMU-TPW 99-10
\vspace*{5ex}
\begin{center}
\Large
{\bf Exact Relativistic Gravitational Field of a Stationary 
	Counterrotating Dust Disk}\\[4ex]
\large C.~Klein\\
{\em Institut f\"ur Theoretische Physik, Universit\"at T\"ubingen,\\
Auf der Morgenstelle 14, 72076 T\"ubingen, Germany}\\[2ex]
O.~Richter\footnote{address after September 1, 1999: Max-Planck-Institut f\"ur
Mathematik in den Naturwissenschaften, Inselstra{\ss}e 22-26, 04103
Leipzig, Germany}\\
{\em Sektion Physik der Universit\"at M\"unchen,\\
Theresienstra{\ss}e 37, 80333 M\"unchen, Germany}\\[4ex]
\end{center}
\vspace{3ex}
\begin{abstract}
Disks of collisionless particles are important models for certain 
galaxies and accretion disks 
in astrophysics. We present here a solution to the stationary 
axisymmetric Einstein equations which represents an infinitesimally thin 
dust disk consisting 
of two streams  of particles circulating with constant angular velocity in 
opposite directions. These streams have the same density distribution but 
their relative density may vary continuously. In the limit of only one 
component of dust, we get the solution for the rigidly rotating dust 
disk previously given by Neugebauer and Meinel, in the limit of identical 
densities, the static disk of Morgan and Morgan is obtained. We 
discuss the 
Newtonian and the ultrarelativistic limit, 
the occurrence of ergospheres, and the regularity of the 
solution.
\end{abstract}
\vspace{3ex}
\centerline{PACS numbers: 04.20.Jb, 02.10.Rn, 02.30.Jr}
\end{titlepage}

\thispagestyle{empty}
\

\newpage
\setcounter{page}{1}

In astrophysics thin disks of collisionless matter, so called dust, are 
discussed as models for certain galaxies (see e.g.\ \cite{binney}) or for 
accretion disks. A fully relativistic treatment of these 
models is necessary if a black hole is present since black holes are genuinely 
relativistic objects. But also the exact treatment of dust disks without 
central object would provide deep insight both in the mathematical 
structure of the field equations and in the physics of rapidly rotating 
relativistic bodies since dust disks can be viewed as a limiting case for 
extended matter sources. Corresponding exact solutions hold 
globally -- in the vacuum and in the matter region -- and can thus 
provide physically realistic testbeds for numerical codes. Since Newtonian 
dust disks are known to be unstable and since there are hints by numerical 
work (see e.g.\ \cite{bawa}) that the same holds in the relativistic 
case, such solutions could be taken as exact initial data for numerical 
collapse calculations. Whereas the Newtonian theory of such disks 
is well established (see \cite{binney} and references given 
therein), the same holds in the relativistic case only for static disks 
which can be interpreted as consisting of two counter-rotating streams of 
matter with vanishing total angular momentum. The first disk of this type 
was considered by Morgan and Morgan \cite{morgan}. Infinitely extended dust 
disks with finite mass were studied by Bi\v{c}\'ak, Lynden-Bell and 
Katz \cite{blk} in the static case and by Bi\v{c}\'ak and Ledvinka  
\cite{ledvinka} in the stationary case. The first to construct the exact 
solution for a finite stationary dust disk were Neugebauer and Meinel 
\cite{neugebauermeinel1} who gave the solution for the rigidly 
rotating dust disk which was first treated numerically by Bardeen  and 
Wagoner \cite{bawa}. They solved the corresponding boundary value problem 
for the  Einstein equations with the help of a corotating coordinate system.

In this letter we present a class of new disk solutions to the 
Ernst equation where such a coordinate system cannot be used. The disks 
of finite radius 
$\rho_0$  consist of two counter-rotating components of dust with 
respective density $\sigma^{\pm}(\rho)$.  The 
angular velocity of both streams of particles is of the same constant  
absolute value $\Omega$ but of different sign. The relative density 
$\gamma=(\sigma^+-\sigma^-)/(\sigma^++\sigma^-)$ is a constant with respect to 
$\rho$ and $\zeta$ which varies 
between one, the rigidly rotating dust disk \cite{neugebauermeinel1}, 
and zero, the static Morgan and Morgan disk \cite{morgan}. Interestingly, 
observations \cite{rubin} indicate that the galaxy NGC 4550 is built from 
two counter-streaming stellar components. We are able to 
give the explicit solution for the above configuration in dependence of the 
three parameters $\rho_0$, $\Omega$ and $\gamma$ which parametrize a Riemann 
surface of 
genus 2. We discuss physically interesting features like the static limit, 
the Newtonian regime and the ultrarelativistic limit. 
We study the occurrence of ergospheres and the absence of 
singularities in the allowed range of the physical parameters. 

\noindent {\em Newtonian dust disks\/}\\
It is instructive to consider first the Newtonian case where the 
gravitational field is given by a scalar potential $U$ which is a solution to 
the Laplace equation $\Delta U=0$.  The potential $U$ has to be everywhere regular 
except at the disk where the balancing of the centrifugal and the 
gravitational force leads to boundary values for $U$. We use dimensionless 
($\rho_0=1$)
cylindrical coordinates ($\rho, \zeta, \phi$) and put the disk in the 
$\zeta=0$-plane which leads at the disk to
\begin{equation}
	U_{\rho}(\rho,0)=\Omega^2 \rho
	\label{newton1}.
\end{equation}
For constant $\Omega$, this can be easily integrated to give 
$U(\rho)=U_0+\frac{1}{2}\Omega^2 \rho^2$ where the constant $U_0$ is 
related in the relativistic case  to the central  redshift.
Notice that there are no effects due to counter-rotation in the Newtonian 
case since $\Omega$ enters (\ref{newton1}) quadratically. The 
solution to this problem can be  
constructed e.g.\ with Riemann-Hilbert techniques (see \cite{jgp2}) 
which lead in the case of the Laplace equation to a potential of the form 
\begin{equation}
U(\rho,\zeta)=-\frac{1}{4\pi 
{\rm i}}\int_{\Gamma}^{}\frac{\ln G(\tau) d\tau}{\sqrt{(\tau-\zeta)^2+\rho^2}}
\label{newton2}
\end{equation}
which is a function on the Riemann surface of genus zero given by $\mu_0^2(\tau)
=(\tau -\zeta)^2+\rho^2$; here $\ln G(\tau)$ is an analytic function and 
$\Gamma$ is the covering of the 
imaginary axis in the upper sheet between $-{\rm i}$ and ${\rm i}$. 
In the Newtonian 
case, the situation is equatorially symmetric which leads at the disk to 
\begin{equation}
	U(\rho,0)=-\frac{1}{2\pi {\rm i}}\int_{0}^{{\rm i}\rho}
	\frac{\ln G(\tau) d\tau}{\sqrt{\tau^2+\rho^2}}
	\label{newton3}.
\end{equation}
This is an Abelian integral equation for $\ln G$ since the left hand side 
is given by the integral of (\ref{newton1}). It has the solution $\ln 
G=4\Omega^2(\tau^2+1)$ where a singular ring at the rim of the disk is excluded 
by the condition $G(\pm{\rm i})=1$. The condition implies that $U$ is continuous 
at the rim, and it fixes the constant $U_0=-\Omega^2$. This shows that the 
solution indeed only depends on the two parameters $\rho_0$ and $\Omega$.

\noindent {\em The relativistic case\/}\\
The metric describing the exterior  (i.e.\ the vacuum region) 
of an axisymmetric, stationary rotating body
can be written in the Weyl--Lewis--Papapetrou form (see~\cite{exac})
\begin{equation}
	{\rm d} s^2 =-{\rm e}^{2U}({\rm d} t+a{\rm d} \phi)^2+{\rm e}^{-2U}\left(
	{\rm e}^{2k}({\rm d} \rho^2+
	{\rm d} \zeta^2)+
	\rho^2{\rm d} \phi^2\right),
	\label{vac1}
\end{equation}
where 
$\partial_{t}$ and $\partial_{\phi}$ are two commuting asymptotically 
timelike respectively spacelike Killing vectors. 
In this case, the vacuum field equations are equivalent 
to the Ernst equation for the potential $f$, 
\begin{equation}
f_{z\bar{z}}+\frac{1}{2(z+\bar{z})}(f_{\bar{z}}+f_z)=\frac{2 }{f+\bar{f}}
f_z f_{\bar{z}}\label{vac10}\enspace,
\end{equation}
where $f={\rm e}^{2U}+{\rm i}b$ 
and the real function $b$ 
is related to the metric functions via
$b_{z}=-({\rm i}/\rho){\rm e}^{4U}a_{z}$.
Here the complex variable $z$ stands for $z=\rho+{\rm i}\zeta$. 
With a solution of the Ernst equation, 
the metric function $U$ follows directly from the definition of the Ernst 
potential whereas $a$ and $k$ can be obtained from $f$ via quadratures.

The complete integrability of this equation allows in principle 
to give explicit solutions for boundary value problems. For
two-dimensionally extended matter sources, e.g.\ 
infinitesimally thin disks, one gets global solutions since one encounters 
ordinary differential equations in the matter region
which provide boundary data for the vacuum 
equations. Riemann-Hilbert problems (see e.g.\ \cite{prd}) generate 
solutions to the Ernst equation with free functions which have to be 
determined by the boundary data. However this is only 
possible uniquely in the case of Cauchy data, which would lead in general to 
singular solutions of the 
elliptic field equations. In \cite{neugebauermeinel1} this problem was 
circumvented by the use of a corotating coordinate system, a technique 
that is not applicable in the case of differential rotation or for several 
matter components. 

In \cite{prd} we have shown that the matrix Riemann-Hilbert problem 
is gauge equivalent to a scalar problem on a Riemann surface 
which can be explicitly solved in the case of rational boundary data, 
i.e.\ belongs to Korotkin's \cite{korot1} finite gap solutions. The 
explicit form of the solution allows in principle to solve 
boundary value problems directly. The branch points of the Riemann 
surface are used as a discrete degree of freedom. Since the solutions 
already have the required
regularity properties (see \cite{prl,prd2}), there can be no problems with 
singularities as in the direct  approach in the matrix case. 
The purely algebro-geometric treatment has the further advantage that
it is not limited by the possibility to introduce special 
coordinate systems.
 
\begin{figure}[t]
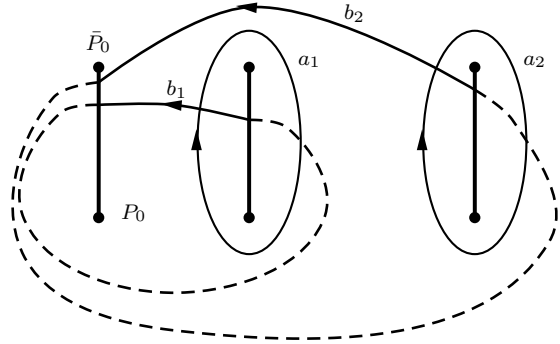

\setlength{\unitlength}{1cm}
\begin{center}
\psset{unit=1cm}
\pspicture[](0,0)(5.5,2)
\psline[linewidth=1.5pt]{*-*}(0,-1)(0,1)
\rput[bl](.3,-1.05){$\scriptstyle P_0$}
\rput[b](0,1.2){$\scriptstyle\bar{P}_0$}
\psellipse[](2,0)(.7,1.5)
\psdots*[dotstyle=triangle*,dotscale=1.2 3](1.3,0)
\rput[b](2.8,1){$\scriptstyle a_1$}
\psline[linewidth=1.5pt]{*-*}(2,-1)(2,1)
\psellipse[](5,0)(.7,1.5)
\psdots*[dotstyle=triangle*,dotscale=1.2 3](4.3,0)
\rput[b](5.8,1){$\scriptstyle a_2$}
\psline[linewidth=1.5pt]{*-*}(5,-1)(5,1)
\pscurve[showpoints=false,linewidth=1pt]{-}(0,.5)(1,.5)(2,.3)
\psdots*[dotstyle=triangle*,dotangle=85,dotscale=1.2 3](1,.5)
\rput[bl](.9,.6){$\scriptstyle b_1$}
\pscurve[showpoints=false,linewidth=1pt,
linestyle=dashed]{-}(2,.3)(2.5,.2)(3,-1)(1,-2)(-1,-1)(-.5,.4)(0,.5)
\pscurve[showpoints=false,linewidth=1pt]{-}(0,.8)(2,1.8)(5,.7)
\psdots*[dotstyle=triangle*,dotangle=85,dotscale=1.2 3](2,1.8)
\rput[b](3.4,1.6){$\scriptstyle b_2$}
\pscurve[showpoints=false,linewidth=1pt,
linestyle=dashed]{-}(5,.7)(5.5,.3)(6,-1.3)(1,-2.5)(-1,-1.5)(-.6,.6)(0,.8)
\endpspicture
\vspace*{2.5cm}
\end{center}
\caption{The homology basis for $\Sigma_2$.}\label{fig1}
\end{figure}

The simplest surface leading to 
equatorially symmetric Ernst potentials is 
of genus $2$ and of
the form $\nu^2(K)=(K-P_0)(K-\bar{P}_0)(K^2-E_1^2)(K^2-\bar{E}^2_1)$ 
where $P_0=-{\rm i}z$ is a moving branch point related to the 
physical coordinates whereas the $z$--independent branch points are given 
by $E_1=-(\alpha_1+{\rm i}\beta_1)$ with positive $\alpha_1$ and $\beta_1$. 
We introduce the standard quantities associated with a Riemann surface, 
with the cut system of figure~\ref{fig1}, namely the 2 differentials of 
the first kind ${\rm d}\omega_i$ normalized by $\oint_{a_i}{\rm d}\omega_j=2\pi{\rm i} 
\delta_{ij}$, the Abel map $\omega_i(P)=\int_{P_0}^{P}{\rm d}\omega_i$ 
which is defined uniquely up to periods,
the Riemann matrix $\Pi$ with the elements $\pi_{ij}=
\oint_{b_i}{\rm d}\omega_j$, 
and the theta function $\Theta(z)=
\sum_{N\in Z^2}^{}\exp\left\{\frac{1}{2}\left\langle\Pi 
N,N
\right\rangle+\left\langle
z,N
\right\rangle\right\}$. The 
normalized (all $a$--periods zero) differentials of the third kind with 
poles in $p$ and $q$ and residues $+1$ respectively $-1$ 
are denoted  by ${\rm d} \omega_{pq}$. Then one can show that 
\begin{equation}
	\bar{f}(\rho,\zeta)=\frac{\Theta(\omega(\infty^{+})+u)}{
	\Theta(\omega(\infty^{-})+u)}
\exp\left\{
\frac{1}{2\pi{\rm i}}\int\limits_\Gamma
\ln G(\tau){\rm d}\omega_{\infty^{+}\infty^-}(\tau)
\right\}
\label{rel1},
\end{equation}
where the two-dimensional vector $u$ has the components
$u_i=\frac{1}{2\pi {\rm i}}\int_{\Gamma}^{}\ln G 
{\rm d}\omega_i$ and $\Gamma$ as in the Newtonian case, is a solution to 
the Ernst equation which is everywhere regular except at the disk if 
$\Theta(\omega(\infty^{-})+u)\neq 0$. We can also give the metric function 
$a$ (see \cite{korot1} and \cite{prd2}) and $k$ (see \cite{korotmat} and 
references given therein)
explicitly. Notice that solutions of the form (\ref{rel1}) are 
characterized by one real valued function $G$ and two real numbers, e.g.\ 
$\alpha$ and $\beta$ defined by $E_1^2=:\alpha+{\rm i}\beta$ where $\beta$ has to 
be positive.

Dust disks within general relativity are best described with the help of 
Israel's covariant formalism \cite{israel} where the regions $\zeta>0$ 
and $\zeta<0$ are matched at the hypersurface $\zeta=0$. The surface 
stress-energy tensor of the disk, for counter-rotating 
dust $S_{ab}=\sigma^+ v_a^+v_b^+ +\sigma^- v_a^-v_b^-$ 
where $v_i^{\pm}$ is the velocity of the respective component of the 
the two-dimensional dust,  is related to the jump of the
extrinsic curvature at the disk. If we put 
$Z=(a+\gamma/\Omega)e^{2U}$, $\lambda=2\Omega^2 e^{-2U_0}$ and 
$\delta=(1-\gamma^2)/\Omega^2$, this leads via the Einstein equations to 
the boundary conditions 
\begin{equation}
	\left(e^{2U}\right)_{\zeta}=\frac{Z^2+\rho^2+\delta e^{4U}}{2\rho 
	Z}b_{\rho},\quad 
	b_{\zeta}=\frac{e^{2U}}{Z}-
	\frac{Z^2+\rho^2+\delta e^{4U}}{2\rho Z}\left(e^{2U}\right)_{\rho}
	\label{rel2}.
\end{equation}
The second equation may be integrated to give $Z^2-\rho^2+\delta 
e^{4U}=\frac{2}{\lambda}e^{2U}$. Notice that $\lambda$ can be viewed as a
parameter that indicates how relativistic the situation is: the larger it is,
the larger is the angular velocity and the central redshift.

We can now state the following theorem which is the main result of this 
letter.\\
{\em Theorem:\/}\\
The solution of the boundary value problem (\ref{rel2}) for the Ernst 
equation is given by an Ernst potential of the form (\ref{rel1}) with
\begin{equation}
	\alpha=-1+\frac{\delta}{2},\quad \beta^2=\frac{1}{\lambda^2}+\delta 
	-\frac{\delta^2}{4}
	\label{rel3}
\end{equation}
and 
\begin{equation}
	G(\tau)=\frac{\sqrt{(\tau^2-\alpha)^2+\beta^2}+\tau^2+1
	}{\sqrt{(\tau^2-\alpha)^2+\beta^2}-\left(\tau^2+1\right)}
	\label{rel4}.
\end{equation}
We briefly sketch the proof which uses in principle only fundamental 
theorems of  algebraic geometry which can 
be found in the standard literature (see e.g.\ \cite{farkas,algebro}). The 
main idea is to establish the relations between the real and the imaginary 
part of the Ernst potential enforced by the given Riemann surface 
independently of $G$\footnote{The simplest example for 
this concept is provided by the function ${\rm e}^{{\rm i}w}$ which is the 
analogue of theta functions on the Riemann surface of genus zero. In this 
case we have of course the simple relation for the real and imaginary part 
that  $\cos^2 w+\sin^2 w=1$ for 
all $w$.}. This selects the class of boundary value problems 
that can be solved on a certain Riemann surface.  We note that the 
steps below can all be performed uniquely what implies that the form of 
the solution is unique at least among genus 2 solutions.\\
{\em Proof:\/}\\
1. The Jacobi inversion theorem ensures that the equations 
$\omega(X)-\omega(D)=u$ where $X=X_1+X_2$ and $D=E_1+(-\bar{E}_1)$ are 
divisors can always be solved for the divisor $X$. The divisor $X$ is 
called non-special if the solution is unique (how to overcome problems with 
special divisors is discussed in \cite{prd}). If we introduce this divisor in 
(\ref{rel1}), we end up with the algebraic formulation of the hyperelliptic 
solutions in \cite{meinelneugebauer} and \cite{korotneu}.\\
2. The reality of the $u_i$ leads to the condition 
$\omega(X)+\omega(\bar{D})=\omega(\bar{X})+\omega(D)$.
Abel's theorem implies that this condition is equivalent to the existence of a rational 
function on $\Sigma_2$ with zeros in $X+\bar{D}$ and poles in $\bar{X}+D$. 
We can thus express $X$ via 
$be^{-2U}$ and a second real quantity as the solution of a system of 
algebraic equations.\\
3. If we introduce the divisor $X$ in the expression for the metric function 
$a$ (\cite{korot1,prd}) we can use so-called root functions (see \cite{algebro}) 
to express $X$ via $a$ and $be^{-2U}$ alone. \\
4. Differentiating with respect to the physical coordinates and using the 
equatorial symmetry in the equatorial plane, we arrive at a set of 
equations which contain only $e^{2U}$, $a$, $b$ and their derivatives. 
Using the boundary conditions (\ref{rel2}) together with the definition 
of $b$, we can show that these equations are identically satisfied.\\
5. The constant $\beta$ is used as an integration constant in the 
integration of the second equation in (\ref{rel2}).\\
6. The function $G$ is determined as in the Newtonian case as the solution 
of an Abelian integral equation, e.g. the equation following from
$\omega_1(X)- \omega_1(D)=u_1$. The regularity condition $G(\pm {\rm i})=1$ then 
fixes $\alpha$.\\

\noindent {\em Discussion:\/}\\
The above disk solutions  have several interesting limiting cases. The 
simplest is of course $\delta=0$, the rigidly rotating dust disk 
\cite{neugebauermeinel1}, here in the form \cite{prl}. 
Mathematically more intriguing is the limit of the Morgan and Morgan disk. 
It was argued in \cite{prl} that the hyperelliptic solutions  of the form 
(\ref{rel1}) are all non-static. This holds of course only in the case of 
regular Riemann surfaces. A real Ernst potential, i.e.\ a solution which 
belongs to the static Weyl class, is only possible on a degenerated 
Riemann surface where the branch points coincide. This is a well-known 
limiting case of algebro-geometric solutions of integrable equations in 
which the soliton solutions are obtained. Since the solutions via 
B\"acklund transformations correspond to the `solitonic' solutions of the 
Ernst equation, Korotkin \cite{korot1} was able to obtain the Kerr 
solution in such a limit.

The problem of the parametrisation of the solution via $\lambda$ and 
$\delta$ which is enforced by the solution process is that the dependence 
on the physical parameters $\Omega$ and $\gamma$ is concealed by a factor
$e^{2U_0}$ which is itself 
a hyperelliptic function of $\delta$ and $\lambda$. In 
the static limit, we get however the simple relation 
$\delta=2(1+\sqrt{1+1/\lambda^2})$ where $\beta=0$. This means that the 
branch points coincide pairwise on the real axis since $\alpha>0$ in this 
case. The Ernst potential thus simplifies to 
\begin{equation}
	U(\rho,\zeta)=-\frac{1}{4\pi {\rm i}}\int_{\Gamma}^{}\frac{\ln 
	G(\tau) d\tau}{\sqrt{(\tau-\zeta)^2+\rho^2}}
	\label{rel5}
\end{equation}
where $G=1-4(\tau^2+1)/\delta$, i.e.\ the expected potential of
the Morgan and Morgan disk in the form \cite{counter}. This disk thus 
provides a nice example where a stationary solution of the Einstein equations 
for an extended body  is linked continuously to a static solution via a 
parameter.

The Newtonian limit of the disks is reached for $\lambda\to0$. In this case, 
we have $G\approx 1$ which implies that the solution approaches Minkowski 
spacetime as expected. Since $\ln G\approx 2\lambda (\tau^2+1)$ 
and $U$ is given by (\ref{newton1}), we get the above Newtonian limit. As 
expected, counter-rotation i.e.\ $\delta$ does not play a role in this case. 
The other extreme limit is the ultrarelativistic limit. The corresponding  
limit of the Morgan and Morgan disk
is of course static. It is reached for $\delta\to4$ or $\lambda\to\infty$ and
describes a disk where the matter streams rotate with the velocity of light
(see e.g.\ \cite{counter}). The density of the matter diverges at the origin 
where the central redshift diverges, too, 
but the mass of the disk remains finite. For the non-static
disks, the exterior of the disk can be interpreted as the extreme Kerr 
solution as in \cite{bawa,prl}. The disk can thus be viewed as being hidden 
behind the horizon of the Kerr metric.  The exact form of these extreme 
configurations will be the subject of further investigation.

In \cite{prl,prd} we have given the necessary conditions for the occurrence 
of ergospheres. Numerically we find that the ergospheres always form first in 
the disk. If more matter is counter-rotating, ergospheres are less likely to 
appear since they are due to gravitomagnetic effects which vanish in the 
static limit of two counter-rotating 
streams with identical density. The necessary condition for an ergosphere to 
reach the rim of the disk is that $\lambda$ takes the value $2/(1-\delta)$ 
which is only possible for $\delta<1$. For larger values of $\delta$, 
ergospheres  thus cannot meet the equatorial plane outside the disk.
In figure 2 we have plotted for given $\delta$ 
the values $\lambda$, $\rho$ of the common points of 
the ergosphere and the disk. For small values of $\lambda$, there will
be no ergospheres since we are close to the Newtonian regime. For increasing
$\lambda$, there will form a ring-like ergosphere at some radius $\rho$. 
If $\lambda$
is increased further, the then roughly toroidal ergosphere hits the disk at two 
values of $\rho$.
The larger value of $\rho$ reaches the rim of the disk at a finite value of 
$\lambda$ for $\delta<0$ or is strictly smaller than 1 for $\delta\geq1$. 
The inner radius of the 
ergosphere will hit the axis in the ultrarelativistic limit for 
$\lambda=\lambda_c$.

\begin{figure}[t]
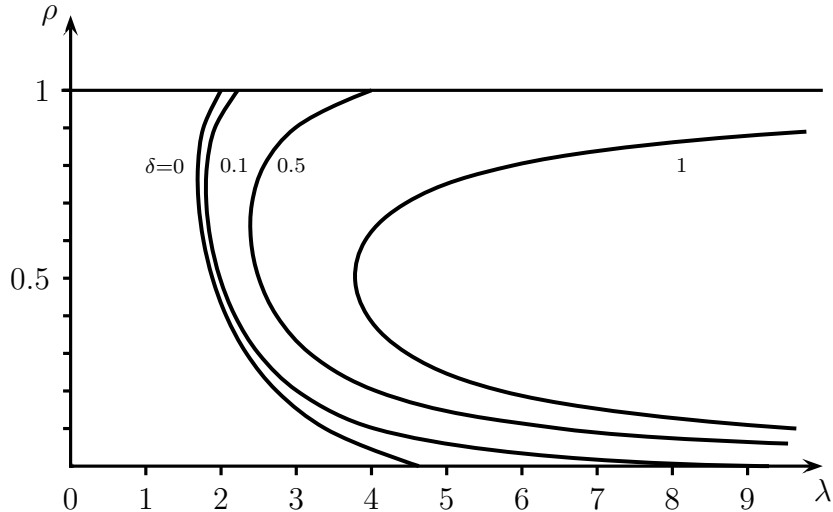

\begin{center}
\psset{xunit=1cm,yunit=5cm}
\pspicture[](0,0)(10,1.2)
\psaxes[linewidth=1.2pt,ticks=all,tickstyle=bottom,labels=x,ticksize=0.1cm]{->}(0,0)(10,1.2)
\multips(0,0)(0,0.1){10}{\psline[linewidth=1.2pt]{-}(-0.1,0.1)(0,0.1)}
\uput*[180](0,1.2){$\rho$}
\uput*[180](-0.1,0.5){0.5}
\uput*[180](-0.1,1){1}
\uput*[270](10,0){$\lambda$}
\uput*[180](1.693801880462834,0.8){$\scriptstyle\delta=0$}
\uput*[0](1.813282004330471,0.8){$\scriptstyle 0.1$}
\uput*[0](2.571764097729977,0.8){$\scriptstyle 0.5$}
\uput*[0](7.879303148300849,0.8){$\scriptstyle 1$}
\psline[linewidth=1.2pt]{-}(0,1)(10,1)
\pscurve[linewidth=1.5pt,linearc=.5]{-}(4.629665061122439,0)(3.385718587568784,0.1)
(2.739927961265721,0.2)(2.337518940461063,0.3)(2.068542375129843,0.4)(1.885672999259529,0.5)
(1.766351901953976,0.6)(1.70154006637709,0.7)(1.693801880462834,0.8)(1.764838078811257,0.9)
(2,1.0)
\pscurve[linewidth=1.5pt,linearc=.5]{-}(9.28477,0)(4.03874,0.1)(3.012520224836469,0.2)
(2.497041584455297,0.3)(2.185105795302819,0.4)(1.985720202150017,0.5)(1.863096122794898,0.6)
(1.804412898982018,0.7)(1.813282004330471,0.8)(1.918125005573852,0.9)(2.222222222,1.0)
\pscurve[linewidth=1.5pt]{-}(9.53796,0.06)(6.506730952937188,0.1)(4.060732058000517,0.2)
(3.180954955536678,0.3)(2.739602170275319,0.4)(2.502811073544734,0.5)(2.398348087461363,0.6)
(2.411624196608576,0.7)(2.571764097729977,0.8)(2.985502363318199,0.9)(4,1.0)
\pscurve[linewidth=1.5pt]{-}(9.64803113758976,0.1)(5.726768168697757,0.2)(4.474136263773968,0.3)
(3.946415727859136,0.4)(3.781317039579762,0.5)(3.915357515755177,0.6)(4.457052736521245,0.7)
(5.879303148300849,0.8)(9.78175,0.89)
\endpspicture
\end{center}
\caption{Ergospheres in the disk in dependence of $\lambda$ and $\delta$.}\label{2}
\end{figure}

It makes little sense to increase the parameter $\lambda$ beyond this 
critical value $\lambda_c$ where the solution still formally exists. 
These regions are most probably non-physical  since 
the solutions may have negative ADM-mass (see \cite{prl}) and have 
singularities. The allowed 
parameter range is thus $0<\lambda<\lambda_c$ and $0\leq \delta \leq 
2(1+\sqrt{1+1/\lambda^2})$.
It can be shown that the regularity condition $\Theta(\omega(\infty^-)+u)\neq0$ 
is always fulfilled in the allowed parameter range. 

Thus we have shown that it is possible by purely algebro-geometric methods 
to construct a solution to the Ernst equation which describes a non-static
counter-rotating dust disk. The fact that this is possible without the use 
of special coordinate systems gives rise to the hope that astrophysically 
interesting objects like dust disks with differential rotation or with a central 
black hole or disks with surface tensions can be constructed in selected 
cases. The open question is whether the resulting algebraic equations can 
still be handled analytically.

\noindent {\em Acknowledgement}\\
We thank J.~Frauendiener, H.~Pfister and U.~Schaudt
for helpful remarks and hints. The work was supported by the DFG.

\end{document}